\documentclass[epj]{svjour}
\usepackage{empheq}
\usepackage{amsmath}
\usepackage{amssymb}
\usepackage{dsfont}

\usepackage{graphicx}% Include figure files
\usepackage{dcolumn}% Align table columns on decimal point
\usepackage{bm}% bold math
\usepackage{textcomp}
\usepackage{wasysym}
\usepackage{stmaryrd}
\usepackage{slashed}
\usepackage{epsfig}
\usepackage{hyperref}
\usepackage{hyperref}
\usepackage{stackrel}
\usepackage{color}
\usepackage{comment}

\def \ee{\end{equation}}
\def \be{\begin{equation}}
\def \eea{\end{eqnarray}}
\def \bea{\begin{eqnarray}}

\begin{document}
\title{The stepwise path integral of the relativistic point particle}
\author{Benjamin Koch\inst{1} \and Enrique Mu\~noz\inst{1}% etc
\thanks{\emph{Present address:} Insert the address here if needed}%
}                     % Do not remove
\offprints{}          % Insert a name or remove this line
\institute{Pontificia Universidad Cat\'olica de Chile \\ Instituto de F\'isica, Pontificia Universidad Cat\'olica de Chile, \\
Casilla 306, Santiago, Chile}
\date{Received: date / Revised version: date}
\abstract{
In this paper we present a stepwise construction of the
path integral over relativistic orbits in Euclidean spacetime.
It is shown that the apparent problems
of this path integral,
like the breakdown of the naive Chapman-Kolmogorov relation,
can be solved by a careful analysis of the overcounting
associated with local and global symmetries.
Based on this, the direct calculation of the quantum propagator of the relativistic point particle in the
path integral formulation results from
a simple and purely geometric construction.
 % end of PACS codes
} %end of abstract
\maketitle

\tableofcontents

%%%%%%%%%%%%%%%%%%%%%%%%%%%%%%%55
\section{Introduction}
%%%%%%%%%%%%%%%%%%%%%%%%%%%%%%%55

Local Lagrangian symmetries and relativity are essential in modern quantum physics.
However, the simplest unification attempt fails dramatically: 
Lagrangian  path integrals of relativistic point particles are considered intractable.
The novelty of this article is that by considering a previously unnoticed symmetry, those problems are overcome
and an exact calculation of the full propagator of the relativistic point particle is achieved.

If one remembers that the relativistic point particle is the simplest system with general covariance,
it becomes clear that
its understanding will be crucial for the consistent formulation of more complex theories with
the same (extended) symmetry such as quantum gravity or string theory.

%%%%%%%%%%%%%%%%%%%%%%%%%%%%%%%55
\subsection{The problem}
%%%%%%%%%%%%%%%%%%%%%%%%%%%%%%%55

Since the first applications in non-relativistic quantum mechanics \cite{Feynman:1950ir},
the path integral formulation of quantum theories has developed
in huge steps towards a quantum field theory of fundamental interactions.
A common consequence of advancing in huge steps
is that one leaves obstacles and possible subtleties unexplored on the way.
This happened with the Path Integral (PI) formulation of the relativistic point particle.
It is the purpose of this paper to close that gap and
to resolve some misconceptions and problems that persisted until today in the context of 
this fundamental topic.

Even though, the PI of the relativistic point particle is the logical
continuation of the non-relativistic PI it has been largely omitted
on the way to quantum field theory.
One reason is that attempts to realize the PI of the relativistic point
particle has presented large complications 
\cite{Teitelboim:1982,Henneaux:1982ma,Redmount:1990mi,Fradkin:1991ci,Padmanabhan:1994}.
Among them, one can distinguish between those that 
seem to be of technical nature and those that seem to be of conceptual nature.

\begin{itemize}
\item[1] Technical complications:\\
At the first sight a technical issue arises from the appearance of non-Gaussian integrals. 
This issue can be
avoided by the use of auxiliary field variables in the Hamiltonian action, which allow to
relate (at least at the classical level) the original action to a quadratic action \cite{Brink:1976,Brink:1977,Fradkin:1991ci},
which is sometimes called einbein formalism. 
Those methods, allow to obtain the expected Klein Gordon propagator from the PI of the relativistic point particle in $D$ dimensions
\be\label{propKG}
K\sim \frac{1}{(k^2+M^2)}.
\ee

As shown in the Appendix \ref{appendA}, a direct
calculation of the PI with the Lagrangian action in $D$ dimensions and without auxiliary fields is 
actually possible.
The problem is however, that
it leads to a  propagator
\be\label{propEnrique}
K^{(n)}\sim \frac{1}{(k^2+M^2)^{n(D+1)/2}}.
\ee

Here, $n$ is the number of intermediate slices.
Apparently (\ref{propEnrique}) does not have the expected form of (\ref{propKG}).
This difference was the first motivation for this study on the direct PI of the relativistic point particle.
\item[2] Conceptual complications:\\
Leaving the technical issues aside, there is a
much more disturbing fact that complicates the understanding of the PI of the relativistic point particle:
The Chapman-Kolmogorov (CK) equation for Markovian processes is not satisfied.
This means that the standard notion of probability is not preserved in the process of free 
relativistic propagation.

As an example, let's show this for  
the propagator (\ref{propKG}) in position space
\be\label{prop0}
K(0,\vec{x})=\mathcal{N}\int d^{D}k \frac{\exp(i\vec{k}\cdot\vec{x})\sqrt{2}M}{k^2+M^2},
\ee
where $\mathcal{N}$ is a normalization constant.
The usual CK condition sates that:
``Propagating from $0$ to $\vec{x}_1$ and then from $\vec{x}_1$ to $\vec{x}_2$ and finally integrating over all $\vec{x}_1$,
must be equivalent
to propagating from $0$ to $\vec{x}_2$''.
Applying this definition to the propagator (\ref{prop0}) gives
\begin{eqnarray}
\label{kol0}
Kol(0,\vec{x}_2)&=&\int d^{D}x_1 K(0,\vec{x}_1)K(\vec{x}_1,\vec{x}_2)\nonumber\\
&=&\mathcal{N}^2 \int d^{D}k \frac{\exp(i\vec{k}\cdot\vec{x}_2)2M^2}{(k^2+M^2)^2}.
\end{eqnarray}
This is of course not the form of the original propagator (\ref{prop0}), which
is the well known and un-understood problem of the path integral of the relativistic point particle.

In the literature there exist different stances on this embarrassing problem.
Mostly, it is just taken as hint that at relativistic velocities the assumption
of a single particle theory breaks down. It is argued that the energy available 
at such velocities would
allow for interactions that again allow for the production of multi particle state~ \cite{Kleinert:book}. 
This argument is however not
very convincing since one was dealing with a free theory without interactions at the first place.
Another stance is to try to fix this problem
by the redefinition of the probability measure~\cite{Jizba:2008,Jizba:2010pi}. 
\end{itemize}

In this paper we propose an elegant solution to those two fundamental problems.
In \cite{Koch:2017bvv} it has been shown that for the case of the relativistic point particle action
it is important to consider a particular local symmetry of the corresponding action.
The preceding paper \cite{Koch:2017bvv} further shows
how formal symmetry considerations
in Minkowski spacetime
in combination with the Fadeev-Popov procedure 
\cite{Faddeev:1967fc} allow to perform
the full functional path integral of the relativistic point particle. 
The idea of this paper is to abstain from abstract technical formulations
and to show instead that the problem can be perfectly understood
and solved by very basic geometric considerations.
Further, several additional explicit calculations and complementary
examples are shown below.

The paper is organized as follows:
The introduction is completed by making notion of the symmetries 
of the relativistic point particle and by a definition of the PI measure taking into account those symmetries.
In section II several Euclidean PIs with one intermediate step are calculated in arbitrary dimensions
and in section III it is proven that those one-step propagators are already the full propagators for
the given theory.
Section IV contains a discussion on the CK relation and the conclusion.
Throughout the paper all formulas and discussions
will be given in the imaginary time formalism corresponding to the Euclidean metric.

%%%%%%%%%%%%%%%%%%%%%%%
\subsection{The relativistic point particle}
%%%%%%%%%%%%%%%%%%%%%%%%%%%%%%%55

The action for a relativistic point particle in $D$ dimensions is
\be\label{actionRPP}
S=\int_{\lambda_i}^{\lambda_f} d\lambda \cdot m \sqrt{\left(\frac{d \vec x}{d\lambda}\right)^2},
\ee
with $\vec x(\lambda_i)=\vec x_i$ and $\vec x(\lambda_f)=\vec x_f$.
This is simply the mass times the geometric length of a given path $\mathcal{P}$.
It is interesting to note that this action in Minkowski space-time can not be equivalent to the action in the
einbein formalism \cite{Brink:1976,Brink:1977,Fradkin:1991ci}, 
for space-like and timelike paths.
This can be seen from the fact that the integration over timelike and space-like paths with (\ref{actionRPP})  
would involve both complex and real values for the action.
Thus, the Euclidean path integral over (\ref{actionRPP}) means probably a restriction on the type of paths
that is allowed in the non-Euclidean version of the path integral.
A detailed analysis of the corresponding integral in Minkowski space-time will be given in \cite{Koch:2018}.

This action and its corresponding Lagrangian are equipped with several symmetries
which will be important for the formulation of a consistent path integral.
\begin{itemize}
\item[({\bf a})] Global Poincar\'e invariance: This can be seen from the fact that the action
is invariant under global rotations and shifts of the coordinate system in $D$ dimensions.
\item[({\bf b})] Local Lorentz invariance: This means that the Lagrangian is invariant under
local rotations in $D$ dimensions of the vector $(d \vec x)/(d\lambda)$ at any point along the trajectory.
A formal argument on why this symmetry, which is not a classical gauge symmetry, is actually important 
in this given context was given in \cite{Koch:2017bvv}.
\item[({\bf c})] Weyl invariance: This means that the Lagrangian does not depend on the way that
$\lambda$ parametrizes a path $\mathcal{P}$. The change to any other function $\tilde \lambda(\lambda)$ would 
leave the Lagrangian invariant.
\end{itemize}
In the following the symmetry ({\bf{a}}) will be used to choose the coordinate system such
that $\vec x_i=0$ and that $\vec x_f$ is different from zero in only one component.
The symmetries ({\bf{b}}) and ({\bf{c}}) are symmetries which have to be
treated with care when it comes to realizing an integral over different paths, since
two seemingly different paths could be actually physically equivalent. 
The over-counting of physically equivalent paths
would result in a wrong weight of some paths with respect to others.

%%%%%%%%%%%%%%%%%%%%%%%%%%%%%%%55
\subsection{General considerations on the explicit form of the PI measure}
%%%%%%%%%%%%%%%%%%%%%%%%%%%%%%%55

%{\color{blue} 
%The ambiguity in the path integral measure is fixed by requiring that the quantum theory has no anomaly. According to Fujikawa's method \cite{Fujikawa}, an anomaly will appear upon quantization if the path integral measure is not invariant under any of the classical symmetries of the action. For the case of the relativistic point particle, one such classical symmetry is scale invariance: $x^\mu\rightarrow \alpha x^\mu$, since $ds^2=g_{\mu\nu}\dot{x}^\mu \dot{x}^\nu$ remains invariant. Thus, the correct path integral measure for $n$ slicings is
%\begin{align}\label{}
%Dx^\mu(\lambda)=\mathcal{N}\ \prod_{i=1}^N \frac{d^Dx_i}{|x_f-x_i|^D}
%\end{align}
%where $\mathcal{N}$ is a numerical, spacetime-independent irrelevant constant.  
%}

\begin{figure}[hbt]
   \centering
\includegraphics[width=9cm]{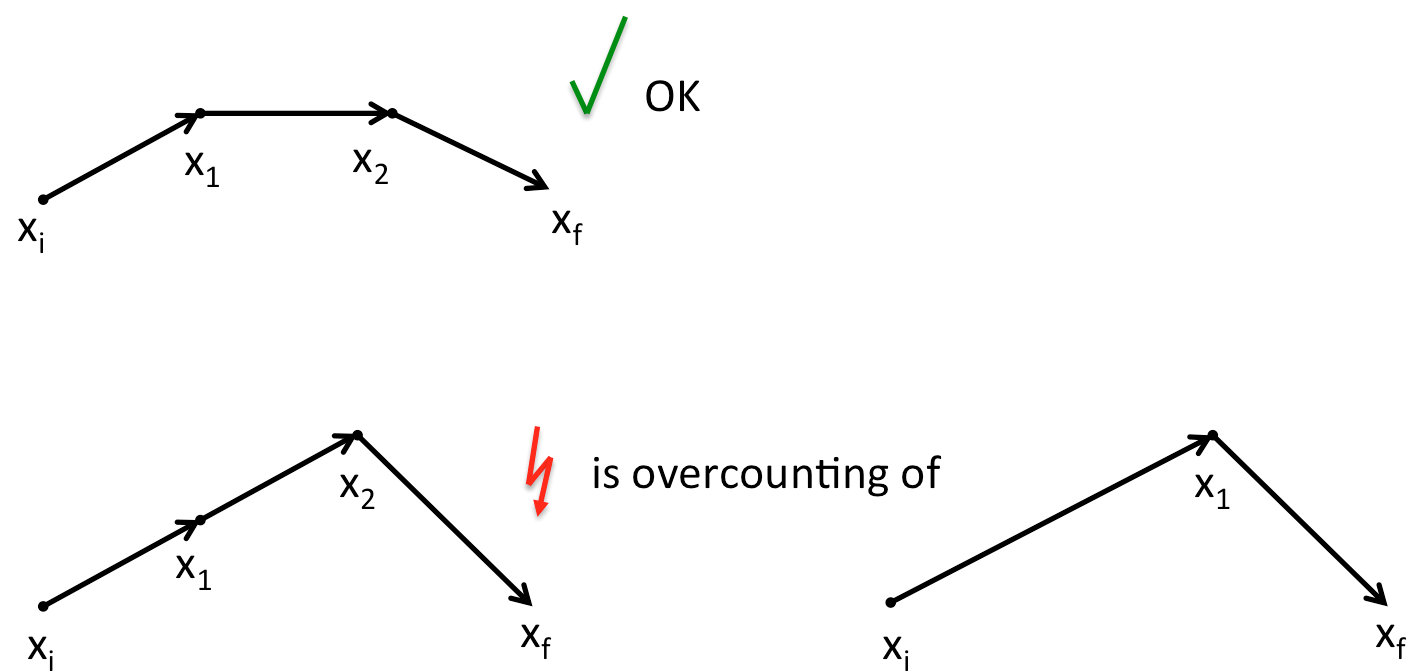}
  \caption{\label{OCfig} Exemplification of possible overcounting of one path
  with two intermediate steps, which is already counted in the PI with one intermediate step. 
  }
\end{figure}

When one formally writes the functional integral ${\mathcal{D}}x(\lambda)$
one has to give a clear definition of this measure.
The common definition of the PI measure is
\bea\label{measure}
K(x_i,x_f)&\equiv&\int_{\vec{x}(t_i)=\vec{x}_i}^{\vec{x}(t_{f})=\vec{x}_f} {\mathcal{D}}x(\lambda)
\exp\left[-\int dt{\mathcal{L}}\right] \\ \nonumber
&=&\lim_{n\rightarrow \infty}{\mathcal{N}}_{D, n}(t)\prod_{j=1}^{n}\int  dx_j^D 
\exp\left[-\sum_j \Delta t{\mathcal{L}}_j\right]  \\ \nonumber
&=&\lim_{n\rightarrow \infty} K^{(n)}(x_i, x_f),
\eea
where the normalization ${\mathcal{N}}_{D, n}(t)$ is a function 
of the external time $t$, the number of time slicings $n$ used, and of the dimensions $D$.
Usually this normalization is fixed from imposing the Kolmogorov relations.
However, there are several issues that arise when one tries to apply this naive measure (\ref{measure})
to the relativistic point particle which are all related to an overcounting of certain paths:

It is easy to see that in the definition (\ref{measure}) one 
is not just counting any possible path between $\vec x_i$ and $\vec x_f$.
One actually counts paths which have straight sections multiple times. Let's exemplify this
for the case of two slicings $n=2$. All configurations where the position $\vec x_1$ is on the classical
path between $\vec x_i\rightarrow \vec x_2$ correspond actually to the same path
$\vec x_i\rightarrow \vec x_2\rightarrow \vec x_f$ as it is shown in figure \ref{OCfig}.
%
%%%%%%%%%%%%%%%%%
%%%%%%%%%%%%%%%%%
Strictly speaking they should only be counted once but according to (\ref{measure}) this path would
be counted multiple times.
Usually, this type of overcounting is completely irrelevant, since the number of paths that
are not overcounted grows much faster with $n$ and $D$ than the number of paths where this overcounting occurs.
In simple words, it is very unlikely that $\vec x_1$ happens to be on the classical
path between $\vec x_i\rightarrow \vec x_2$.
However, in problems with local symmetries this might not necessarily be the case.
In order to avoid overcounting of identical paths right from the start, 
one can improve 
(\ref{measure}) by stating for $K=K(x_i,x_f)$
\bea\label{measure2}
K&\equiv& \lim_{n\rightarrow \infty} \sum_{j=1}^n K^{(j)}(x_i,x_f)\\ \nonumber
&\equiv&\left.
\lim_{n\rightarrow \infty} \sum_{j=1}^n {\mathcal{N}}_{ D, j}(t) \prod_{l=1}^j \int  dx_l^D\right|_{NOC}
\exp\left[-\sum_j \Delta t{\mathcal{L}}_j\right] \\ \nonumber
&=&{\mathcal{N}}_{D, 1}(t) \left.\int dx^D_1\right|_{NOC}
\exp\left[-t/2({\mathcal{L}}_{i,1}+{\mathcal{L}}_{1,f})\right]+ \\ \nonumber
&&{\mathcal{N}}_{D, 2}(t) \left.\int dx^D_1\int dx^D_2\right|_{NOC}
\exp\left[-t/3({\mathcal{L}}_{i,1}+\dots)\right]\nonumber\\
&&+\dots,
\eea
where $|_{NOC}$ stands ``integrate but without overcountig'' in the sense described before.
For example, when performing the two slicing integral one has to avoid overcounting of configurations
which are equivalent to one slicing paths and when doing higher number of slicings ($j=n$)
one has to avoid all the configurations which are actually already counted in ($j<n$) paths.
Even thought this definition is morally superior to (\ref{measure}), it is in most cases highly impractical
since one has to take into account numerous conditions when actually performing the integrals.
Fortunately, in most cases, the term with the highest number of integrals $j=n$ represents a $D+n$ dimensional volume
which dominates the sub-leading contribution which is a $D+n-1$ dimensional volume and the definition
(\ref{measure2}) is equivalent to (\ref{measure}).\\

Even though, the overcounting issue seems to be settled by the definition
(\ref{measure2}), one has to be careful, since imposing a $|_{NOC}$ condition
means in most practical cases a fixing of the symmetry. 
Different fixings can correspond to different restrictions of the measure,
which would generate unphysical anomalies  \cite{Fujikawa:1979ay}. However, we expect
that the symmetries of the action are also symmetries of the measure.
This can be solved by redefining the measure with a multiplicative 
factor $\Delta_i$ such that it is invariant under different configurations within the symmetry
%
%\begin{widetext}
\bea\label{measure3}
K&\equiv& \left.\nonumber
\lim_{n\rightarrow \infty} \sum_{j=1}^n {\mathcal{N}}_{ D, j}(t)
 \prod_{l=1}^j \int  dx_l^D\right|_{NOC} \Delta_l
e^{\left[-\sum_r t_{r+1,r}{\mathcal{L}}_{r+1,r}\right]} \\ \nonumber
&=&{\mathcal{N}}_{D, 1}(t) \left.\int dx^D_1\right|_{NOC}
\Delta_1
\exp\left[-t_{1,i}{\mathcal{L}}_{i,1}-t_{1,f}{\mathcal{L}}_{1,f})\right] \\ 
&&+{\mathcal{N}}_{D, 2}(t) \left.\int dx^D_1\int dx^D_2\right|_{NOC}
\Delta_2
\exp\left[\dots\right] 
+\dots.
\eea
%\end{widetext}
The external normalization factors ${\mathcal{N}}_{D, j}(t)$ are fixed by a dimensional argument.

Relation (\ref{measure3}) is the definition of the measure ${\mathcal{D}}x$,
that will now be used for the PI of the relativistic point particle.

%%%%%%%%%%%%%%%%%%%%%%%
\section{Euclidean path integrals with one intermediate step}
%%%%%%%%%%%%%%%%%%%%%%%

In this section we will discuss how
the considerations on the measure, overcounting,
and symmetries, are applied to the one-slicing propagator $K^{(1)}$.
This will be explicitly done in one and two dimensions, before it
is generalized to $D$ dimensions.
Before actually turning to the relativistic point particle it is instructive to discuss (\ref{measure3}),
in particular the meaning of the restriction $|_{NOC}$, for the case
of non-relativistic quantum mechanics.

%%%%%%%%%%%%%%%%%%%%%%%%%%%%%%%55
\subsection{The non-relativistic path integral in two dimensions}
%%%%%%%%%%%%%%%%%%%%%%%%%%%%%%%55

\begin{figure}[hbt]
   \centering
\includegraphics[width=9cm]{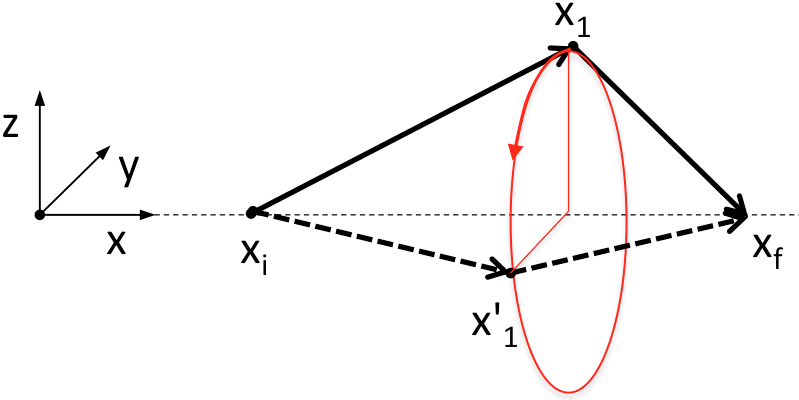}
  \caption{\label{NRfig} Non-relativistic PI in three dimensions 
  with equivalent intermediate
  points $\vec x_1$.
  }
\end{figure}

Some of the redundancies that will be important for the
relativistic case are already present in the path integral of the non-relativistic
point particle. 
It is therefore instructive to discuss this case first and to show that
in the non-relativistic case the over-counting over equivalent
paths does not do any harm since it can be absorbed into a constant normalization factor.
For the classical Lagrangian $\vec v^2/(2M)$
one can construct the propagator with one intermediate step by composing
two propagations with time lapse $\Delta t/2$ each $K^{(1)}(\vec x_i, \vec x_f)=K^{(1)}$ reads
\be\label{k1NR}
K^{(1)}=\int d^{D} x_1 \cdot {\mathcal{N}}_{1} 
\exp \left[ -\frac{M}{2}\left(\frac{(\vec x_1-\vec x_i)^2}{\Delta t/2}+\frac{(\vec x_f-\vec x_1)^2}{\Delta t/2}\right)\right].
\ee
There are two Lagrangians and one action in this problem which are
\bea\nonumber
{\mathcal{L}}_{i1}&=&2M \left(\frac{(\vec x_1-\vec x_i)^2}{\Delta t^2}\right),\\ \nonumber
{\mathcal{L}}_{1f}&=&2M \left(\frac{(\vec x_f-\vec x_1)^2}{\Delta t^2}\right), \;{\mbox{and}}\\ \label{Svonr}
S&=& \Delta t \left({\mathcal{L}}_{i1}+{\mathcal{L}}_{1f} \right).
\eea
Now one wants to see whether there are local transformations
which allow to apply a continuous change of $\vec x_1$ while leaving ${\mathcal{L}}_{i1}$ and ${\mathcal{L}}_{1f}$ invariant.
For $D=2$ one finds that those two conditions completely fix the values of $\vec x_1$ as long as 
$\vec x_i \neq \vec x_f$.
Thus, there is no overcounting in the propagator in two dimensions as long as $\vec x_i \neq \vec x_f$.
At first sight it seems that with identical initial and final positions, there is
an overcounting, even for $D=2$. But this is actually not the case, since
seemingly equivalent paths become distinguishable, when seen from a different
Galilelian reference system. Thus, for non-relativistic paths with two spatial dimensions,
there is no overcounting.
However, when $D\ge 3$, there is an additional freedom in the orientation
of the vector $\vec x_i+\vec x_1$ with respect to $\vec x_i$.
This situation is shown in figure~\ref{NRfig}.
%
%%%%%%%%%%%%%%%%%
%%%%%%%%%%%%%%%%%
This additional freedom corresponds to paths which are all redundant
since they do not change any of the terms in~(\ref{Svonr}).
Thus, due to the $|_{NOC}$ condition in (\ref{measure3}), those paths should
not be included when integrating.
The question that arises is, does this $|_{NOC}$ condition affect the usual
propagator in non-relativistic quantum mechanics?

In order to see this redundant part of the integral more clearly
one can perform a series of coordinate transformations.
First one shifts the integration variables $\vec x'=\vec x- (\vec x_f -\vec x_i)/2$,
then one goes to radial coordinates in three dimensions $\{x',y',z'\}\rightarrow \{r,\theta,\alpha \}$
with the Jacobian $dx' dy' dz' \rightarrow dr \sin(\theta) d\theta d\alpha$, and finally
one changes the radial integration for an integration over the action $dr \rightarrow dS (dr/dS)$ according to (\ref{Svonr}).
Integrating over $\theta$ gives
\bea \label{k1NR2}
K^{(1)}(\vec x_i, \vec x_f)&=& \left(\int_0^{2\pi} d\alpha \right) \frac{(\Delta t)^{3/2}}{(2 M)^{3/2}}\\ \nonumber
&&\cdot \int_{S_c}^\infty dS \exp [-S] 
\sqrt{S-S_c}\cdot 
{\mathcal{N}}_{1}.
\eea
Thus,
\be \label{k1NR2}
K^{(1)}(\vec x_i, \vec x_f)=\left(\int_0^{2\pi} d\alpha \right) \frac{(\Delta t)^{3/2}}{(2 M)^{3/2}}\frac{\sqrt{\pi}}{2}\exp [-S_c] {\mathcal{N}}_{1},
\ee
where $ \mathcal{N}_1$ is assumed to be position  independent and where $S_c=\frac{M}{2}\frac{(\vec x_f-\vec x_i)^2}{\Delta t}$.
One observes from (\ref{k1NR2}) that the volume corresponding to the overcounting of local rotations $d\alpha$ factors out nicely. 
Note that this coordinate transformation corresponds to the one-slicing version of the Fadeev-Popov procedure.
Finally, one sees from this relation that one can choose a simple normalization
\be\label{NormNR}
{\mathcal{N}}_{1}=\frac{2(2 M)^{3/2}}{\left(\int_0^{2\pi} d\alpha \right) \sqrt{\pi} \Delta t^{3/2}},
\ee
such that the classical propagator $K^{(0)}\sim \exp (-S_{c})$ fulfills the Kolmogorov relation
\be\label{KolNR}
K^{(0)}(\vec x_i, \vec x_f)=\int dx_1 dy_1 dz_1 K^{(0)}(\vec x_i, \vec x_1) K^{(0)}(\vec x_1, \vec x_f).
\ee
The important point of this discussion is that instead of integrating over the equivalent configurations
one could actually fix the angle $\alpha$ to some value which corresponds to dividing (\ref{k1NR2})
by a factor of $2\pi$. When one does this fixing, 
one also has to exclude those redundant configurations
in the non-relativistic Kolmogorov relation (\ref{KolNR}).

However, since all this symmetry fixing at the end of the day results in a simple rescaling of the 
normalization (\ref{NormNR}) by a factor of $2\pi$, one sees that considering the 
symmetry ({\bf{b}})
actually leaves all results of non-relativistic quantum mechanics unchanged.
The situation is different for the relativistic path integral, as shown in the following subsections.

%%%%%%%%%%%%%%%%%%%%%%%
\subsection{The relativistic path integral in one dimension}
%%%%%%%%%%%%%%%%%%%%%%%

Let's start the discussion of the relativistic PI with the most
simple case, paths in one dimension.
Already this case shows some non-trivial features
since even though there is no global or local Lorentz symmetry,
the Weyl symmetry~({\bf{c}}) is already present in this case.
As usual,
one can fix this symmetry by choosing a unit length
evolution parameter such that for each path 
\be\label{lambdaFix}
\left(\frac{d \vec x}{d\lambda}\right)^2=1,
\ee
for which the Lagrangian reads
\be\label{LR}
 {\mathcal{L}}=M.
\ee

The fundamental building block of the path integral is
the exponential of the classical action times some normalization constant
which we will choose equal to $M$ for the one dimensional case
\be\label{K01D}
K^{(0)}(x_i, x_f)= M \cdot \exp (-M|x_f-x_i|).
\ee
The next step towards constructing the relativistic path integral in one dimension
consists in introducing one intermediate slice $x_1$.
According to (\ref{measure3}), the one slicing propagator $K^{(1)}(x_i, x_f)=K^{(1)}$ would then be
\begin{eqnarray}
K^{(1)}&=&{\mathcal{N}}_{1, 1}(t_{i,f})\int_{-\infty}^{+\infty}dx_1|_{NOC} \Delta_1 \cdot \nonumber\\
&&K^{(0)}(x_i, x_1)\cdot K^{(0)}(x_1, x_f).
\end{eqnarray}
The normalization and the anomaly cancelation in one dimension is trivial 
${\mathcal{N}}_{1, 1}(t)=1=\Delta_1$. 
Before integrating, one has still to take into account
the redundancy explained in figure \ref{OCfig}.
Whenever $x_1$ is between $x_i$ and $x_f$,
the two-step propagation $x_i\rightarrow x_1 \rightarrow x_f$ is 
physically exactly the same path as the direct connection
 $x_i \rightarrow x_f$ and should not be counted over and over again.
Thus, according to (\ref{measure3}) the right one-step propagator  in one dimension 
$K^{(1)}(x_i, x_f)|_{NOC}=\tilde K^{(1)}$
is
\bea\label{K11Da}
\tilde K^{(1)}&=&\int_{-\infty}^{x_i}dx_1 K^{(0)}(x_i, x_1)\cdot K^{(0)}(x_1, x_f)\\ \nonumber
&&+
\int_{x_f}^{+\infty}dx_1 K^{(0)}(x_i, x_1)\cdot K^{(0)}(x_1, x_f),
\eea
where we assumed $x_f>x_i$.
A trivial integration gives
\be\label{K11D}
K^{(1)}(x_i, x_f)|_{NOC}= M \cdot \exp (-M|x_f-x_i|) = K^{(0)}(x_i, x_f),
\ee
which means that for one dimension the initial expression (\ref{K01D}) 
is equal to the one slicing propagator (\ref{K11D}).
This has important consequences for the generalization
to $n$ intermediate steps, as it will be discussed in a later section.

Remember that due to the fact that the one dimensional case does not
have the local symmetry {\bf{b})} it was possible to waive all length dependent
normalization factors defined in (\ref{measure3}) and choose ${\mathcal{N}}({t, 1, 1})=1=\Delta_1$.
For the two dimensional case, the symmetry {\bf{b})} is present and
thus one has to consider those non-trivial contributions.

%{\color{red}Comment: At least for this 1D example, it seems that the NOC measure could be simply expressed by
%$dx_1|_{NOC} \equiv dx_1\Theta(x_i - x_1)\Theta(x_1 - x_f)$. Is it possible to further generalize this simpler definition
%to $D > 1$?}
%{\color{blue} Maybe, I do not know???}

%%%%%%%%%%%%%%%%%%%%%%%
\subsection{The relativistic path integral in two dimensions}
%%%%%%%%%%%%%%%%%%%%%%%

\begin{figure}[hbt]
   \centering
\includegraphics[width=10cm]{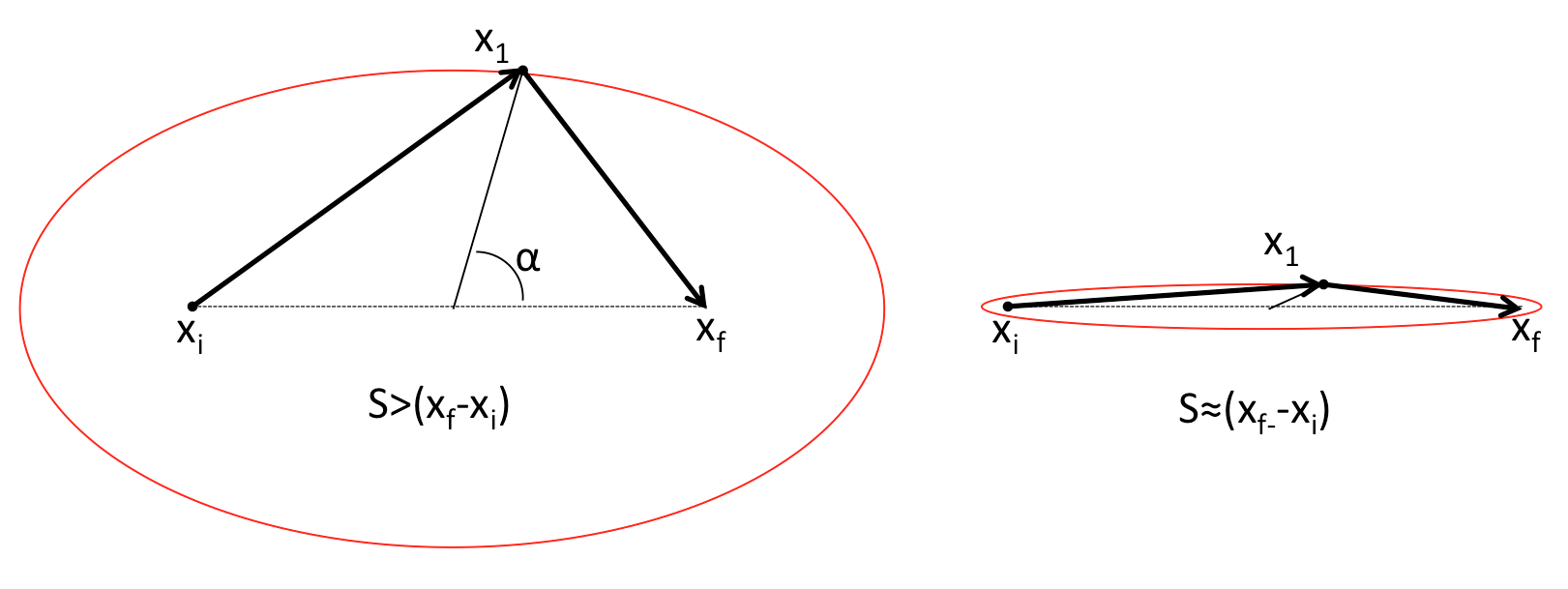}
  \caption{\label{R2} Left: Relativistic PI in two dimensions for paths with same action $S$, where the red curve
  represents all  equivalent points $x_1$ for a given action value $S>S_{cl}$.
Right:
  It is interesting to note that in the classical limit, when $S\rightarrow S_{cl}$,
  the elliptic contour of equivalent points $x_1$ collapses to the classical path.
  This means that the two local symmetries ({\bf{b}}) and ({\bf{c}}) are indistinguishable at the
  classical level.}
\end{figure}

It is instructive to continue
the discussion of the problem in two dimensions with one intermediate step.
According to (\ref{measure3}), the right one-slice propagator $K^{(1)}(\vec{x}_i,\vec{x}_f)=K^{(1)}$ is 
\bea\label{OneSDdim}
K^{(1)}&=&{\mathcal{N}}_{ D, 1}(t_{i,f}) \int d^Dx_1|_{NOC} \Delta_1 \cdot \\ \nonumber &&
\exp{[-S_0(\vec{x}_i,\vec{x}_1)]}
\cdot 
\exp{[-S_0(\vec{x}_1,\vec{x}_f))]}.
\eea
In two Euclidean dimensions ($\vec x=(x(\lambda), y(\lambda))$), the relativistic path integral~(\ref{OneSDdim}) reads
\bea\label{R1t}
K^{(1)}&=&
{\mathcal{N}}_{ 2, 1}(t_{i,f})
\int d^2x_1 |_{NOC} \Delta_1 \cdot \\ \nonumber &&
\exp{\left[-M|\vec x_1-\vec x_i|-M|\vec x_f-\vec x_1|\right]}.
\eea
For simplicity let's choose the initial position as origin of the Cartesian coordinate system
$\vec x_i =0$. 
As part of this integral one can consider paths that 
all contribute with the same action to this path integral.
One finds that the surface of
points with equal action  is shown 
by the ellipse in figure \ref{R2}.
This curve is generated by all different paths ($\vec x_i\rightarrow \vec x_1\rightarrow \vec x_f$) with the same total length.
%
%%%%%%%%%%%%%%%%%
%%%%%%%%%%%%%%%%%
%
However,
the important feature of the relativistic case is that
the Lagrangian~(\ref{LR}) is the same for every single point
along any of those paths with $S=const.$ and $\vec x_i,\, \vec x_f$ fixed.
This is a reflection of the fact that
local Lorentz symmetry ({\bf{b}}) has not been fixed.
Thus, if one naively 
counts all the points on the relativistic elliptic contour in figure~\ref{R2}, one is actually 
counting paths which are connected by a symmetry transformation.
Instead, one should count only one point out of the elliptic contour
by properly fixing the freedom introduced by the symmetry ({\bf{b}}).

In order to cast the integral (\ref{R1t}) in the equal-action form
\bea\label{K12DS}
K^{(1)}(0,\vec x_f)
&=&\int_{S_0}^{\infty} dS \exp (-S) \cdot \Omega_{R,2}^{(1)}(S; 0,\vec x_f),
\eea
one can transform the Cartesian coordinates $x_1, y_1$ into slightly modified elliptical coordinates $\chi_i=(S,\, \alpha)$, with
\bea\label{elliptical}
x_1&=& \frac{S}{2M} \cos (\alpha),\\
y_1&=& \frac{x_f}{2}\sqrt{\left(\frac{S}{x_fM}\right)^2-1}\cdot  \sin (\alpha).
\eea
Here, we have chosen the middle of the classical path $0 \rightarrow \vec x_f$
as the origin of the elliptical coordinate system and $\alpha$ as the angle between
the $x$ axes and the vector $\vec x_1-(\vec x_f-\vec x_i)$, as shown in figure \ref{R2}.
Further, for the Cartesian choice $\vec x_i=0$, $x_f=|\vec x_f- 0|$ stands for the length of the minimal classical
path, while $S$ stands for the length of the quantum path $0 \rightarrow \vec x_1 \rightarrow \vec x_f$. 
The Jacobian of the coordinate transformation
is 
\be\label{detn}
\det
\left(\frac{x_i}{\chi_j}
\right)=\frac{2 \left(S/M\right)^2-x_f^2 (1+\cos (2 \alpha))}{8 \sqrt{\left(S\right)^2-(x_f M)^2}}.
\ee
Thus the one-step 
propagator $K^{(1)}(0,\vec x_f)=K^{(1)}$ is
\bea
K^{(1)}&=& {\mathcal{N}}_{ 2, 1}(t_{i,f})\int_{x_f M}^\infty dS \int_0^{2 \pi} d\alpha \cdot \\ \nonumber 
&& \Delta_1
\frac{2 \left(S/M\right)^2-x_f^2 (1+\cos (2 \alpha))}{8 \sqrt{(S)^2-(x_f M)^2}}
\exp{[-S]}.
\eea
One notes that
the angular integral, which represents the redundancy, does
not simply factorize like in the non-relativistic case.
This anomalous angular dependence in the numerator of (\ref{detn})
comes due to a naively defined measure $dx_1 dy_1$ and has
to be canceled by a proper definition of the multiplicative factor
\be\label{r2}
\Delta_1^{-1}\equiv |\vec x_1-0| \cdot |\vec x_f-\vec x_1|=\frac{1}{8}
\left(2 \left(S/M\right)^2-x_f^2 (1+\cos (2 \alpha))\right)
\ee
which is the simplest non-fractional term containing $\alpha$.
A dimensional analysis shows that the measure ${\mathcal{N}}_{ 2, 1}(t_{i,f})={\mathcal{N}}_{ 2, 1}$ 
in two dimensions is actually independent of $x_{f}\equiv|\vec x_f - 0|$.
With this, one finds
\bea
K^{(1)}(0,\vec x_f)&=&{\mathcal{N}}_{ 2, 1}\left(\int_0^{2 \pi} d\alpha \right) \\ \nonumber &&
\cdot \int_{x_fM}^\infty dS 
\frac{1}{\sqrt{(S)^2-(x_f M)^2}} \exp{[-S]}\nonumber \\ 
&=& {\mathcal{N}}\cdot (2 \pi)  K_0(x_f M).\label{BesselK}
\eea
Please note that the redundant volume $\left(\int_0^{2 \pi} d\alpha \right)$
can now either be eliminated by fixing the angle $\alpha$ or by integrating over the
angle and absorbing the factor in the normalization of the propagator.
A two dimensional Fourier transformation of (\ref{BesselK})
returns again the expected propagator of the free scalar field in Fourier space.

By comparing the one slicing propagator $K^{(1)}(x_f)=(2 \pi) K_0(x_f M)$ with 
the  zero slicing propagator $K^{(0)}(x_f)\sim \exp(- M |x_f|)$,
one notes that in contrast to the one dimensional case, the Kolmogorov relation is not fulfilled when going from zero to one slicing.
This makes it necessary to study a higher number of intermediate steps, which will be done
after discussing the one-step case in $D$ dimensions.

%%%%%%%%%%%%%%%%%%%%%%%
\subsection{The relativistic path integral in $D$ dimensions}
%%%%%%%%%%%%%%%%%%%%%%%

It is straight forward to repeat the discussion of the two dimensional case
for the path integral of the relativistic point particle in higher dimensions. 
When changing to the modified elliptical coordinates in $D$ dimensions $\chi_i=(S,\alpha_1,...,\alpha_{D-1})$
one notes that for one slicing the Jacobian of the transformation
takes the form
\bea
\det
\left(\frac{x_i}{\chi_j}
\right)&=&
\left[2 \left(S/M\right)^2-x_f^2 (1+\cos (2 \alpha_1))\right] g(\alpha_1) \\ \nonumber &&
\cdot\left(S^2-(x_f M)^2\right)^{(D-3)/2}\cdot f(\alpha_2, \dots \alpha_{D-1}).
\eea
This determinant can be written as a product of three functions,
one with the angle $\alpha_1$, 
one with the angles $\alpha_i|_{i>1}$,
and of one function without angles.
Here, $g(\alpha_1)$ and $f(\alpha_2, \dots \alpha_{N-1})$ are the typical
angular dependencies in $D$-dimensional spherical coordinates.
Just like in the two dimensional case, the anomalous non-factorization of the angle
$\alpha_1$, can be corrected by an appropriate
choice of $\Delta_1$.
An anomaly free symmetry fixing choice is again assured for the definition~(\ref{r2}),
independent of the dimension $D$.
The remaining angular functions $g(\alpha_1)\cdot f(\alpha_2, \dots \alpha_{D-1})$
do not mix with $S$ and give simply a solid angle.
Further, in order to compensate the change in dimensionality induced
by each additional spatial integral one has to choose the normalization
with an inverse dimensional factor of $(|\vec x_f - 0|)$
\be\label{NormD}
{\mathcal{N}}_{ D, 1}(|\vec x_f - 0|)\equiv {\mathcal{N}}\cdot |\vec x_f - 0|^{2-D}.
\ee
Note that the $x_f$ dependence of the normalization (\ref{NormD}) can also be obtained
from imposing a matching to the non-relativistic limit of the resulting propagator.
With this normalization and after integrating the factorized solid angle,
the propagator reads
\bea\label{PropDD00}
K^{(1)}&=& \frac{ {\mathcal{N}}}{|x_f|^{D-2}} \int_{x_f M}^\infty dS \exp[-S] 
\left(S^2-(x_f M)^2\right)^{(D-3)/2}.\nonumber\\
\eea
By integrating over $S$ one gets the
one-slicing propagator in $D$ dimensions
\bea\label{PropDD}
K^{(1)}(x_f)
&=&{\mathcal{N}}'' \left(\frac{1}{x_f M}\right)^{D/2-1} K_{D/2-1}(M|x_f|),
\eea
where $K_{D/2-1}$ is the modified Bessel function.

%%%%%%%%%%%%%%%%%%%%%%%
\section{Euclidean path integrals with $n$ intermediate steps}
%%%%%%%%%%%%%%%%%%%%%%%
\label{sec_Nsteps}

Up to now we have calculated the relativistic propagator with one intermediate
step $K^{(1)}_D$.
However, according to (\ref{measure3})
one still has to calculate infinitely many propagators with $n$ intermediate steps $K^{(n)}_D$ and than one has to sum them all up, taking again into account that no overcounting occurs. 
This sounds like a lot of work, unless one has some convenient relation (or theorem) that allows
to deduce all $K^{(n)}_D$ and their sum just from the knowledge of $K^{(1)}_D$.
In non-relativistic quantum mechanics, this powerful tool of simplification
is given in terms of the Kolmogorov relation~(\ref{KolNR}).
It will now be shown, that  there exists a generalization of this relation
for the relativistic PI in one dimension and
that for the relativistic PI in higher dimensions there exists a ``nothing new theorem'' which also
allows to deduce $K^{(n)}_D$ from the knowledge of $K^{(1)}_D$.

%%%%%%%%%%%%%%%%%%%%%%%
\subsection{One dimensional case}
%%%%%%%%%%%%%%%%%%%%%%%

By taking into account the ``no overcounting'' condition
 it was previously shown that, 
 the one-slicing propagator~(\ref{K11D}) 
 has actually the same functional form as the  
 fundamental infinitesimal propagator (\ref{K01D}).
Thus, by iterating this process one obtains that the $n$ step propagator $K^{(n)}(x_i, x_f)$ is also equal to
(\ref{K01D}) and that this propagator does fulfill the
CK relation, as long as one avoids overcounting
in the summation over intermediate steps $x_1$.
Thus, summing all $K^{(n)}$ still gives
\be
K_1(x_f-x_i)\sim K_1^{(n)}(x_f-x_i)= K_1^{(0)}(x_f-x_i),
\ee
up to a normalization constant.

It is instructive to study the one-dimensional propagator in its Fourier representation.
In order to get a relation in Fourier space one can operate
on both sides of (\ref{K11Da}) with $\int_{-\infty}^{\infty} dx_f \exp(+i k x_f)$ which gives after
a shift of integration variables $\tilde x_f=x_f-x_1$ on the right hand side
\be\label{KolF1}
\frac{2 M}{k^2+M^2}=M\frac{2 M}{k^2+M^2}\cdot \frac{2 M}{k^2+M^2}- 2 M\frac{M^2-k^2}{(k^2+M^2)^2}.
\ee
This is the relativistic CK relation in Fourier space in one dimension. 
The unusual but essential piece is the subtraction of the $2 M\frac{M^2-k^2}{(k^2+M^2)^2}$ term
which comes from the missing piece ($M \int_{0}^{x_f}dx_1 \exp (-|x_f|)=M|x_f| \exp (-|x_f|)$ ) 
on the right hand side in (\ref{K11Da}). 
The physical meaning of this subtraction is that in a naive $\int_{-\infty}^{\infty} dx_1$ integration, 
the part from 0 to $x_f$ corresponds
to over-counting of physically equivalent paths.
Thus, the right hand side of (\ref{KolF1}) means
that one can indeed combine two propagators $\sim 1/(k^2+M^2)$ such that they give
again the same form of a propagator $\sim 1/(k^2+M^2)$, 
if one correctly subracts the overcounting part $\sim(M^2-k^2)/(k^2+M^2)^2$.

%%%%%%%%%%%%%%%%%%%%%%%
\subsection{Two dimensional case}
%%%%%%%%%%%%%%%%%%%%%%%

The anomaly free relativistic propagator
in two dimensions with two intermediate steps without overcounting is according to the definition (\ref{measure3}) given by
\bea\label{R2t}
&&K^{(2)}(\vec x_i,\vec x_f)|_{NOC}={\mathcal{N}} \int d^2x_1 d^2x_2|_{NOC}\cdot \Delta_2\nonumber\\
&& \cdot 
\exp{\left[-M(|\vec x_1-\vec x_i|+|\vec x_2-\vec x_1|+|\vec x_f-\vec x_2|)\right]}.
\eea
It is straight forward to see that the anomaly cancelation is provided by
\be
\Delta_2^{-1}=|\vec x_1-\vec x_i| \cdot |\vec x_2-\vec x_1|\cdot |\vec x_f-\vec x_2|
\ee
and that in general, for $n$ intermediate steps
\be
\Delta_n^{-1}=\prod_{j=1}^{n+1} |\vec x_j-\vec x_{j-1}|.
\ee

\begin{figure}[hbt]
   \centering
\includegraphics[width=9cm]{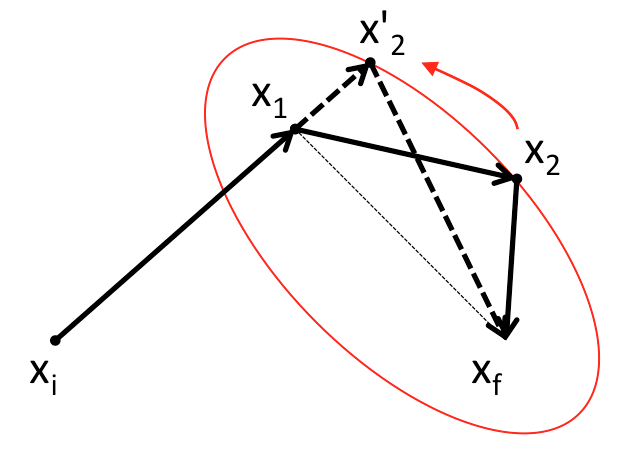}
  \caption{\label{R3} Relativistic PI with two intermediate steps.}
\end{figure}

The situation of two intermediate steps is shown in figure \ref{R3}.
%
%%%%%%%%%%%%%%%%%
%%%%%%%%%%%%%%%%%
%
We now give a geometrical proof that the symmetry-fixed two-step propagator (\ref{R2t}) is given by the one-step
propagator~(\ref{BesselK}). This proof is to be understood within the ``No-Over-Counting'' definition of
the path integral measure (\ref{measure3}).

For any two points $\vec x_1$ and $\vec x_2$ in (\ref{R2t}) one can distinguish
three cases:
\begin{itemize}
\item[1)] If $\vec x_1$ is on the direct connection between $\vec x_i$ and $\vec x_2$, the
step $\vec x_1$ does not contribute a new path to (\ref{R2t}) and those
paths do not contribute to $\int dx_1^2 \int dx_2^2 |_{NOC}$.
\item[2)] If $\vec x_2$ is on the direct connection between $\vec x_1$ and $\vec x_f$, the
step $\vec x_2$ does not contribute a new path to (\ref{R2t}) and those
paths do not contribute to $\int dx_1^2 \int dx_2^2 |_{NOC}$.
\item[3)] If none of the above cases applies, one knows that the path 
$\vec x_1 \rightarrow \vec x_2 \rightarrow \vec x_f$ is equivalent (symmetry {\bf{b}}) to the path 
$\vec x_1 \rightarrow \vec x_2' \rightarrow \vec x_f$ as indicated by the  dashed lines in figure \ref{R3},
where $\vec x_2'$ is chosen such that $\vec x_1$ lies on the direct line between $\vec x_i$ and $\vec x_2'$.
As already shown, this different choice of $\alpha_2$ does not change the measure
contributed by this path.
Thus, the total path $\vec x_i \rightarrow \vec x_1 \rightarrow \vec x_2 \rightarrow \vec x_f$
is equivalent to the total path $\vec x_i \rightarrow \vec x_1 \rightarrow \vec x_2' \rightarrow \vec x_f$.
Since $\vec x_1$ lies on the direct line between $\vec x_i$ and $\vec x_2'$, one falls back to scenario
1).
\end{itemize}
Combining the outcome of those three possible scenarios 
one sees  that  the integration $\int dx_1^2 \int dx_2^2 |_{NOC}=\emptyset$.
It only contains paths which are already present in the one-step integration  $\int dx_1|_{NOC}$ .
Thus one has shown that
\bea\label{K2eqK1}\nonumber
K^{(2)}(\vec x_i,\vec x_f)&=&K^{(1)}(\vec x_i,\vec x_f)|_{NOC}+K^{(2)}(\vec x_i,\vec x_f)|_{NOC}\\ \nonumber
&=&K^{(1)}(\vec x_i,\vec x_f)|_{NOC} + 0 \\ 
&=&{\mathcal{N}} \cdot K_0(x_f),
\eea
where the factor of $2 \pi$ was absorbed in the normalization.
Having shown that two intermediate steps are equivalent
to one intermediate step one can repeat this
procedure with $n$ steps and will always find that 
the propagator is given by (\ref{BesselK}).
This also means that adding intermediate steps to the calculation
of a propagator does not alter (\ref{BesselK}) and thus, the Kolmogorov
relation holds trivially since there are no physically new intermediate points one can add.
In anciant words:
``There is nothing new under the sun''.

%%%%%%%%%%%%%%%%%%%%%%%
\subsection{D dimensional case}
%%%%%%%%%%%%%%%%%%%%%%%

Generalizing (\ref{PropDD}) to $n$ intermediate steps in $D$ dimensions is straight forward.
The relation (\ref{K2eqK1}) holds also in this case, because one
can set the angles $(\alpha_{i,1},...,\alpha_{i,N-1})$ such that new intermediate steps
are on a straight line with the one-intermediate step case.
Thus, the ``nothing new theorem'' holds also in $D\ge 2$ dimensions and
the Kolmogorov relation for the relativistic point particle in $D$ dimensions becomes trivial.

%%%%%%%%%%%%%%%%%%%%%%%
\subsection{A check: Three dimensional non-relativistic PI with two intermediate steps}
\begin{figure}[hbt]
   \centering
\includegraphics[width=9cm]{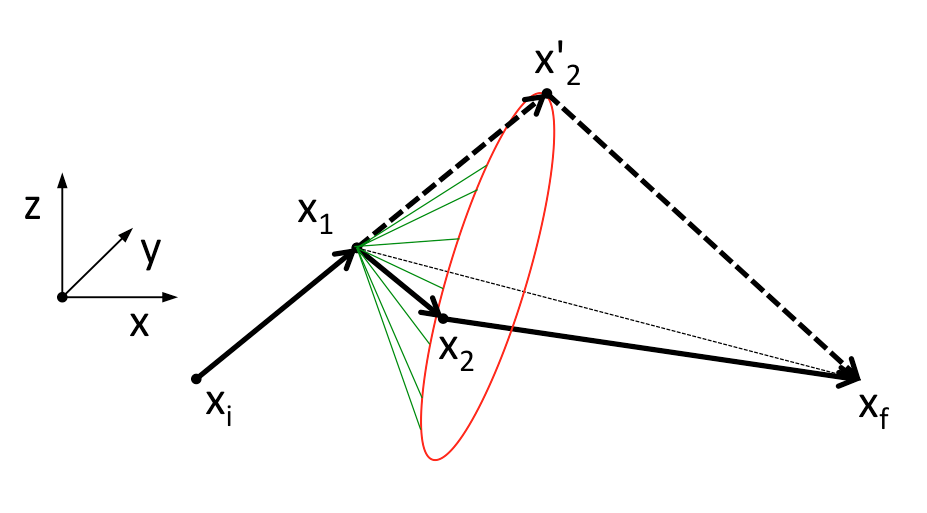}
  \caption{\label{NR3} Non-Relativistic PI in three dimensions with two intermediate steps.}
\end{figure}

%%%%%%%%%%%%%%%%%%%%%%%
Now, since it was shown that the entire relativistic path integral
can be reduced to the one slicing case, one should revisit 
whether something similar happens
in  the non-relativistic case. 
Let's consider a non-relativistic PI in three dimensions
with two intermediate steps $\vec x_i \rightarrow \vec x_1 \rightarrow \vec x_2 \rightarrow \vec x_f$
as shown in figure \ref{NR3}.
%
%%%%%%%%%%%%%%%%%
%%%%%%%%%%%%%%%%%
%
Let's further consider the plane defined by the three points $\vec x_i, \vec x_1, \vec x_f$ and discuss
the different scenarios that arise, when one chooses the second intermediate step $\vec x_2$.
\begin{itemize}
\item[1)] If the point $\vec x_2$ lies in the plane defined by $\vec x_i, \vec x_1, \vec x_f$, in some cases
a direct overcounting can happen.
This occurs when $\vec x_2$ lies on the continuation of the line $\vec x_i \rightarrow \vec x_1$ 
(indicated by the red dashed line in figure \ref{NR3}), this configuration is actually an overcounting,
since it is already considered by a one-slicing path $\vec x_i \rightarrow \vec x_2 \rightarrow \vec x_f$.
This can happen, but it is actually numerically irrelevant since it corresponds to a one dimensional subset of the
three dimensional volume $d^3x_2$.
\item[2)] The same argument holds if $\vec x_1$ lies on the continuation of $\vec x_f \rightarrow \vec x_2$.
\item[3)] If the point $\vec x_2$ lies outside of the plane defined by $\vec x_i, \vec x_1, \vec x_f$ it still can happen
that it is connected to an overcounting by a local rotation. Due to the rotational invariance explained in figure \ref{NRfig}
one can always choose the arbitrary angle $\alpha_2$ such that the transformed point $\vec x_2'$ lies in the plane defined
by $\vec x_i, \vec x_1, \vec x_f$. If this point $\vec x_2'$ lies on the continuation of $\vec x_i \rightarrow \vec x_1$ one
has an overcounting, otherwise not. 
All points $\vec x_2$, which correspond to an overcounting, lie on a cone who's tip is the point $\vec x_1$, who's symmetry axis
is defined by the line $\vec x_1 \rightarrow \vec x_f$, and who's opening is given by the continuation of the line $\vec x_i \rightarrow \vec x_1$.
This configuration is shown by the green surface in figure \ref{NR3}.
Again the volume of this overcounting is two dimensional which is negligible with respect to the three dimensional volume
of $d^3x_2$.
\end{itemize}
As a result of those three scenarios, one can conclude that there is some overcounting in the higher dimensional
non-relativistic case with more than one slicing, but this overcounting is irrelevant since it is of lower dimension than the actual
integral $d^3x_2$. 
Thus, in the non relativistic case one has to actually construct the complete PI by the use of the Kolmogorov relation.
This is in contrast to the relativistic case, where the entire volume $d^Dx_2$ turned out to be
an overcounting.
A generalization of this observation to higher number of dimensions and higher number of slicings
is straight forward.

%%%%%%%%%%%%%%%%%%%%%%%%%%%%%%%55
\section{Discussion and conclusion}
%%%%%%%%%%%%%%%%%%%%%%%%%%%%%%%55

%%%%%%%%%%%%%%%%%%%%%%%%%%%%%%%55
\subsection{The Kolmogorov relation}
%%%%%%%%%%%%%%%%%%%%%%%%%%%%%%%55

It is interesting to discuss the results
(\ref{PropDD}) and the D dimensional generalization of (\ref{K2eqK1})
in the context of the CK relation.

The usual way to state this relation in the non relativistic case is based
on the fact that the one-step propagator $K_{NR}^{(1)}(x_i, x_f)$ takes the same form
as the infinitesimal propagator $K_{NR}^{(0)}(x_i, x_f)$
\be\label{Kol00}
K_{NR}^{(1)}(x_i, x_f)=K_{NR}^{(0)}(x_i, x_f),
\ee
where,
\be
K_{NR}^{(1)}(x_i, x_f)=\int d^Dx K_{NR}^{(0)}(x_i, x)K_{NR}^{(0)}(x, x_f).
\ee
This allows to
construct the complete propagator $K(x_i, x_f)$ by iteration.
Thus, (\ref{Kol00}) is also fulfilled by
the complete propagator 
\be\label{Kol}
K_{NR}(x_i, x_f)=\int d^Dx K_{NR}(x_i, x)K_{NR}(x, x_f).
\ee

In the relativistic case the infinitesimal propagator
$K^{(0)}(x_i, x_f)$ differs from the one-step propagator $K^{(1)}(x_i, x_f)$
\be
K^{(1)}(x_i, x_f)\neq K^{(0)}(x_i, x_f),
\ee
where
\be\label{Kol00R}
K^{(1)}(x_i, x_f)=\int d^Dx  \Delta_x \cdot  K^{(0)}(x_i, x)K^{(0)}(x, x_f)
\ee
and it seems hopeless to construct a PI from an iterative relation.
As already seen, the rescue comes when one takes into account the issue
of overcounting over symmetry-equivalent intermediate steps
when gluing together additional infinitesimal propagators.
By virtue of (\ref{K2eqK1}) one finds that the iteration actually converges
at one intermediate step, since all additional intermediate steps
can be identified as symmetry-equivalent to one intermediate step
\be\label{KolNrel}
K(x_i, x_f)\equiv K^{(n)}(x_i, x_f)=K^{(n-1)}(x_i, x_f) \quad {\mbox{for all}}\quad n\ge 2.
\ee
This is the building brick
 of the relativistic Kolmogorov relation (\ref{KolNrel}) in analogy to the
non-relativistic relation~(\ref{Kol00}).
The naive KG relation fails for the relativistic case since it ignores this issue of overcounting,
and thus it is dominated by a summation over equivalent paths, which
generates the ``Tur Tur effect'' \cite{Ende:1995}  of increasingly deformed
propagators as one continues to introduce intermediate steps.

%%%%%%%%%%%%%%%%%%%%%%%%%%%%%%%55
\subsection{Conclusion}
%%%%%%%%%%%%%%%%%%%%%%%%%%%%%%%55

The usual path integral formulation of relativistic quantum mechanics 
suffers from deep technical and conceptual problems.
A first problem appears when one uses the Lagrangian action and calculates the propagator
in a straight forward way.
It turns out that such a
calculation of the propagator gives a wrong result (see Appendix A).
Even though, one can to some degree circumvent this particular problem
by applying  techniques such as
renormalization procedures combined with
semi-classical approximations \cite{Padmanabhan:1994}, or the introduction of 
auxiliary fields for the Hamiltonian action~\cite{Brink:1976,Brink:1977,Fradkin:1991ci},
there remains a much deeper conceptual problem.
This second problem is that it is not understood how the combination of short time
propagators $K^{(0)}$ gives long time propagators $K^{(n)}$ of the same functional form, in the spirit of
the Chapman-Kolmogorov relation.
This failure is typically taken as hint that it is impossible to
consistently formulate the PI of the relativistic point particle and that 
one has to turn to QFT instead
(at least if one does not want to redefine the usual notion of probability~\cite{Jizba:2008,Jizba:2010pi}).

Those two long standing issues are resolved in this paper.
For this purpose, we make notice of the symmetries ({\bf a}), ({\bf b}), and ({\bf c}) which are present in
this problem.
Then, the usual PI measure is defined in a precise and explicit
way~(\ref{measure3}), taking into account the issue of overcounting of equivalent and identical paths.
Based on this, the relativistic one slicing propagator is calculated in 
a very simple and geometric way (\ref{PropDD}).
The crucial step of the paper was then to show that it is indeed possible
to consistently relate the n-slicing propagator to the one-slicing propagator
by proving the relation~(\ref{K2eqK1}).
Thus, the one-step propagator (\ref{PropDD}) already gives the right result for the full propagator.
This proof makes again heavy use of the symmetries and overcounting conditions discussed before.
The main part of this paper concludes with a discussion on the Chapman-Kolmogorov relation and a
conjecture on other quantum theories with general covariance.

It is hoped and believed that the presented work allows to finally
reconcile the quantum mechanical PI formulation with the
straight forward notion of a relativistic action (\ref{actionRPP}).

%%%%%%%%%%%%%%%%%%%%%%%%%
\section*{Acknowledgements}
We want to thank I.A. Reyes and A. Faraggi for helpful discussions.
B.K. was supported by Fondecyt No 1161150,
E.M was supported by Fondecyt No 1141146.

\appendix

\section{Propagator, direct calculation ignoring issue of overcounting}
\label{appendA}

In this appendix, we present the derivation of Eq.(1). Let us consider 
the D-dimensional relativistic propagator  between $\vec{x}_i  \equiv \vec{x}_0 = 0$ and $\vec{x}_f \equiv \vec{x}_{n+1}$ fixed, for $n$-time slices, and for notational simplicity we adopt Euclidean metric.
\begin{eqnarray}
K(0,\vec{x}_f) &=& \left[\prod_{j=1}^{n}\int d^{D}x_j\right] K(0,\vec{x}_1) \ldots K(\vec{x}_{n},\vec{x}_f)\nonumber\\
&=& \left[\prod_{j=1}^{n}\int d^{D}x_j\right] e^{ -M \sum_{l=0}^{n} |\vec{x}_{l+1} - \vec{x}_{l}| }.
\label{eq_prop1}
\end{eqnarray}
Now, let us liberate the last point $\vec{x}_{n+1}$, by introducing a constraint via the identity $1 = \int d^{D}x_{n+1} \delta(\vec{x}_{n+1} - \vec{x}_f)$, as follows
\begin{eqnarray}
K(0,\vec{x}_f) &=& \left[\prod_{j=1}^{n+1}\int d^{D}x_j\right] e^{ -M \sum_{l=0}^{n} |\vec{x}_{l+1} - \vec{x}_{l}|}\times\delta(\vec{x}_{n+1} - \vec{x}_f)\nonumber\\
&=& \int \frac{d^D k}{(2\pi)^D} e^{-i \vec{k}\cdot \vec{x}_f} K_{n+1}(k,M).
\label{eq_prop2}
\end{eqnarray}
Thus, the integrand in the above expression is nothing else but the Fourier transform of the propagator after $n$
time slices, 
thus corresponding to the multi-dimensional integral
\begin{eqnarray}
K_{N+1}(k,M) &=& \left[\prod_{j=1}^{n+1}\int d^{D} x_{n}\right]\nonumber\\
&&\times e^{-M \sum_{l=0}^{n}|\vec{x}_{l+1} - \vec{x}_{l}| + i \vec{k}\cdot \vec{x}_{n+1}}.
\label{eq_A1}
\end{eqnarray}
We begin by defining the new set of variables
$\vec{y}_0 = \vec{x}_1$,
$\vec{y}_1 = \vec{x}_2 - \vec{x}_1,\ldots$, 
$\vec{y}_n = \vec{x}_{n+1} - \vec{x}_{n}$.
This set of equations can be inverted to yield
\begin{eqnarray}
\vec{x}_1 &=& \vec{y}_0\nonumber\\
\vec{x}_2 &=& \vec{y}_0 + \vec{y}_1\nonumber\\
\vdots\nonumber\\
\vec{x}_{n+1} &=& \vec{y}_0 + \vec{y}_1 + \vec{y}_2 + \ldots \vec{y}_n.
\label{eq_A3}
\end{eqnarray} 
The Jacobian of this transformation is 1.
Therefore, Eq.(\ref{eq_A1}) can be expressed as
\begin{eqnarray}
K_{n+1}(k,M)&=& \left[\prod_{j=0}^{n}\int d^{D}y_j\right] e^{-\sum_{l=0}^{n}(M |\vec{y}_l| + i \vec{k}\cdot \vec{y}_l)}\nonumber\\
&=& \left[
K(k,M)
\right]^{n+1}.
\label{eq_A4}
\end{eqnarray}
Here, $K(k,M)$ represents the Fourier transform of the zero-time-slice propagator, and thus can be expressed in D-dimensional spherical coordinates: $\rho = |\vec{y}_{n}|$,
$d^{D}y_{n} = d\omega_{D} \rho^{D-1} d\rho$,
\begin{eqnarray}
K(k,M) &=& \int d \omega_D \int_{0}^{\infty} d\rho\, \rho^{D-1}\,e^{-M \rho - i\, k \rho \cos\theta}.
\label{eq_A5}
\end{eqnarray}
Here, we use the recurrence property for the differential solid angle $d\omega_D = (\sin\theta)^{D-2} d\theta d\omega_{D-1}$. Considering the integral representation of the Bessel function of the first kind,
\begin{eqnarray}
J_\nu(z) = \frac{(z/2)^\nu}{\Gamma(\nu + 1/2)\Gamma(1/2)}\int_{0}^{\pi} e^{\pm iz\cos\theta}(\sin\theta)^{2\nu} d\theta,
\label{eq_A6}
\end{eqnarray}
the integral in Eq.(\ref{eq_A5}) can be expressed as
\begin{eqnarray}
K(k,M) = \frac{(2\pi)^{D/2}}{(k)^{D/2-1}}\int_{0}^{\infty}d\rho\,\rho^{D/2}\,e^{-M\rho}\,J_{D/2-1}(k\rho).
\label{eq_A7}
\end{eqnarray}
Substituting the expression for the solid angle in $D-1$-dimensions, $\omega_{D-1} = 2 \pi^{D/2-1/2}/\Gamma(D/2-1/2)$, and changing the variable of integration to $x = M \rho$, we have
\begin{eqnarray}
K(k,M) &=& (2\pi)^{D/2} M^{-(D/2+1)} (k)^{1-D/2}\nonumber\\
&&\times \int_{0}^{\infty} e^{-x}\,x^{D/2}\,J_{D/2-1}(k\, x/M).
\label{eq_A8}
\end{eqnarray}
The last integral is obtained analytically, thus giving
\begin{eqnarray}
\int_{0}^{\infty}dx && e^{-x}\, x^{D/2}\,J_{D/2-1}(k\,x/M) = \frac{2^{D/2}}{\sqrt{\pi}}\Gamma(D/2+1/2) \nonumber\\
&&\times
(k/M)^{D/2-1} \left( 1 + (k/M)^{2}\right)^{-D/2-1/2}.
\label{eq_A9}
\end{eqnarray}
Substituting into Eq.(\ref{eq_A8}), we finally have
\begin{eqnarray}
K(k,M) = \frac{(4\pi)^{D/2}}{\sqrt{\pi}}\frac{M\, \Gamma(D/2 + 1/2) }{\left( k^2 + M^2 \right)^{(D+1)/2}}.
\label{eq_A10}
\end{eqnarray}
Inserting this result into (\ref{eq_A4}) one obtains the result stated in (\ref{propEnrique}).

%%%%%%%%%%%%%%%%%%%%%%%%%%%%%%%%%%%%%%%%%%%%%%%%%%
%%%%%%%%%%%%%%%%%%%%%%%%%%%%%%%%

\end{document}